\documentclass[prl,showpacs,floatfix,superscriptaddress,12pt]{revtex4-1}

\usepackage{microtype} 
\usepackage[T1]{fontenc}       
\usepackage{amsmath}
\usepackage{dcolumn}
\usepackage[small]{caption}
\usepackage[english]{babel}
\usepackage{graphicx}
\usepackage{varioref}
\usepackage{epstopdf}
\usepackage{color}
\usepackage{eucal}
\usepackage{float}
\usepackage{soul}
\usepackage{physics}
\usepackage{bbm}
\graphicspath{{./pictures/}}
\usepackage{booktabs}

\begin{document}

\title{Boltzmann Transport in Nanostructures as a Friction Effect}
\author{Andrea Cepellotti}
\affiliation{Theory and Simulations of Materials (THEOS) and National Centre for Computational Design
and Discovery of Novel Materials (MARVEL),
\'Ecole Polytechnique F\'ed\'erale de Lausanne, 1015 Lausanne, Switzerland}
\author{Nicola Marzari}
\email{nicola.marzari@epfl.ch}
\affiliation{Theory and Simulations of Materials (THEOS) and National Centre for Computational Design
and Discovery of Novel Materials (MARVEL),
\'Ecole Polytechnique F\'ed\'erale de Lausanne, 1015 Lausanne, Switzerland}

\begin{abstract}
Surface scattering is the key limiting factor to thermal transport in dielectric crystals as the length scales are reduced or when temperature is lowered.
To explain this phenomenon, it is commonly assumed that the mean free paths of heat carriers are bound by the crystal size and that thermal conductivity is reduced in a manner proportional to such mean free paths.
We show here that these conclusions rely on simplifying assumptions and approximated transport models.
Instead, starting from the linearized Boltzmann transport equation in the relaxon basis, we show how the problem can be reduced to a set of decoupled linear differential equations.
Then, the heat flow can be interpreted as a hydrodynamic phenomenon, with the relaxon gas being slowed down in proximity of a surface by friction effects, similar to the flux of a viscous fluid in a pipe.
As an example, we study a ribbon and a trench of monolayer molybdenum disulphide, describing the procedure to reconstruct the temperature and thermal conductivity profile in the sample interior and showing how to estimate the effect of nanostructuring.
The approach is general and could be extended to other transport carriers, such as electrons, or extended to materials of higher dimensionality and to different geometries, such as thin films.
\end{abstract}

\maketitle

While most physical models address bulk crystals, many properties are strongly altered by the presence of surfaces.
In particular, thermal conductivity in dielectric crystals can be strongly affected by the finite size of a crystal.
As temperature is lowered, or as the material size or its polycrystalline texture reaches micrometer scales, a strong reduction of thermal conductivity is expected, with the result that surface engineering is considered one of the leading strategies for optimizing thermoelectric devices \cite{thermoelectrics:review:snyder,thermoelectrics:review:dresselhaus,cahill:review}.

The current explanation of the mechanism behind the reduction of thermal conductivity by a surface comes from pioneering works of the late 1930s  \cite{deHaas,casimir}.
Broadly speaking, surface effects dominate transport dynamics when the system sizes approach the length-scale of the phonon mean free paths (MFPs), i.e. the distance traveled by phonons between scattering events.
Since a phonon cannot travel unperturbed for a distance larger than the size of the host material, the MFPs can be scaled down by the material size and, being approximately proportional to the phonon MFPs, thermal conductivity will decrease as well.

Despite its appealing simplicity, this picture is built on the approximate hypothesis that phonons and their MFPs are the relevant quantities to characterize thermal conductivity.
This is a correct assumption within the relaxation-time approximation, a simplified description of the scattering dynamics that has nevertheless some important shortcomings.
Notably, it cannot describe heat transport at low temperatures, or in 2D materials even at room temperature, given that under these conditions the scattering events are dominated by normal momentum-conserving processes.
Several works have tried to improve the description of surface scattering; we here briefly mention the contributions of Ziman \cite{ziman} and more recently, Refs. \citenum{majumdar}, \citenum{chen2001} and \citenum{minnich2015}, to name a few: the whole literature on the topic is too vast to be reviewed here.
However, most have tackled the problem within the relaxation-time approximation for the phonon scattering, or by simplifying the phonon diffusion operator \cite{mingo2005}.
Recently, we have shown \cite{relaxons}, starting from the linearized phonon Boltzmann transport equation (LBTE), that an exact description can be achieved by introducing a new heat carrier, called relaxon.
Thermal conductivity can be obtained - without approximating the scattering operator - in terms of relaxons' heat capacity, velocities and MFPs.
However, in Ref. \citenum{relaxons} relaxons have been studied only in the bulk; thus not much is known about surface scattering.
Moreover, relaxons' MFPs differ from those of phonons by about 2 orders of magnitude (e.g. in graphene or in silicon at 300K), and therefore it's not even clear if the standard interpretation mentioned above is applicable.

In this work, we aim to characterize thermal conductivity of a finite-sized crystal by solving exactly a space-dependent LBTE.
We show that the commonly accepted interpretation of surface effects, i.e. that a surface limits the MFP of heat carriers, doesn't hold for the relaxons.
Instead, the heat flux of the relaxon gas is affected by friction coefficients in proximity of surfaces and reaches bulk conditions at large enough distance from the surface.
The theory is applied to a monolayer of MoS$_2$, where the LBTE is built with harmonic and anharmonic interatomic force constants calculated from first principles.
We will focus the examples on two geometries of a 2D material as shown in Fig. \ref{fig1}: (a) a ribbon of infinite length and finite width $W$; and (b) a trench of infinite width and finite length $L$.
The temperature gradient is positioned along the material's length and coincides with the $x$ coordinate axis, while the $y$ axis is located along the crystal width.
Choosing the cartesian origin in the center of mass of the system, the ribbon (trench) surfaces are located at $y=\pm W/2$ ($x=\pm L/2$).


\begin{figure}
\centering
\includegraphics{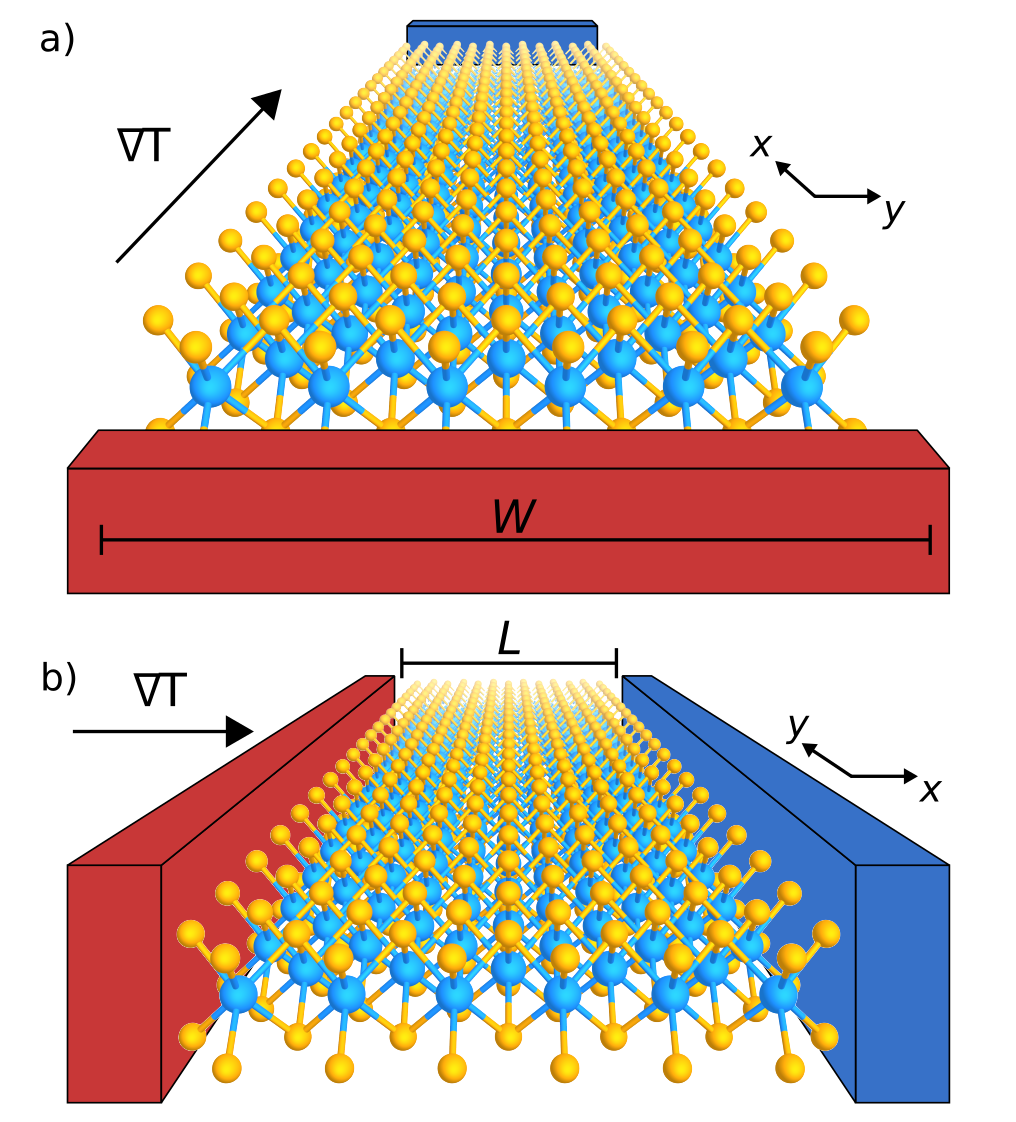}
\caption{Geometries of a MoS$_2$ monolayer considered in this work: (a) a ribbon of infinite length and finite width $W$; and (b) a trench of infinite width and finite length $L$. In the former case, heat sink and source are at infinite distance, while in the latter case they are characterized by an infinite width.}
\label{fig1}
\end{figure}

The main quest of semiclassical transport theories is to determine the space-dependent phonon excitation number $n_{\mu}(x,y)$, where $\mu=(\vec{q},s)$ is a short-hand notation for all possible states ($\vec{q}$ labels the phonon wave-vector and $s$ the phonon branch).
At equilibrium, $n_{\mu}$ loses the space dependence and reduces to the Bose--Einstein distribution function $\bar{n}_{\mu}=1 / (\exp (\hbar \omega_{\mu}/k_BT)-1)$, where $T$ is the temperature and $\omega_{\mu}$ the phonon frequency.
When a thermal gradient $\vec{\nabla} T$ is applied, $n_{\mu}$ deviates from equilibrium by a quantity $\Delta n_{\mu}(x,y)=n_{\mu}(x,y) - \bar{n}_{\mu}$.
To determine $\Delta n_{\mu}$, one solves the phonon LBTE \cite{carruther}:
\begin{equation}
\vec{v}_{\mu} \cdot \vec{\nabla} T \frac{\partial \bar{n}_{\mu}}{\partial T} 
+ \vec{v}_{\mu} \cdot \vec{\nabla} \Delta n_\mu(x,y)
= - \frac{1}{\mathcal{V}} \sum_{\mu'} \Omega_{\mu\mu'} \Delta n_{\mu'}(x,y) \; ,
\label{eq1}
\end{equation}
where $\vec{v}_{\mu}$ is the phonon group velocity and $\mathcal{V}$ is a normalization volume. The scattering matrix $\Omega$ contains in general the rates for all possible phonon transitions $\mu \to \mu'$, computed with the aid of Fermi's golden rule.
Here, we will consider only 3-phonon and phonon-isotope scattering (see Refs. \citenum{relaxons,fugalloPRB} for details of the construction of this matrix).
After solving the LBTE, one obtains the phonon response $\Delta n_{\mu}(x,y)$ and the heat flux \cite{hardy-flux} from $\vec{Q}(x,y) = \frac{1}{\mathcal{V}} \sum_{\mu} \Delta n_{\mu}(x,y) \hbar \omega_{\mu} \vec{v}_{\mu}$.
Finally, thermal conductivity is easily obtained through its definition $\vec{Q} = - k \vec{\nabla} T$.
Note that the dependence of the phonon population on the position inside the crystal propagates to the thermal conductivity, which therefore stops being a bulk property (we will later see under which condition a bulk thermal conductivity can still be defined).

The LBTE can be greatly simplified exploiting symmetries.
For example, a ribbon geometry should be translationally invariant along $x$: the $x$ dependence is lost and, further noting that $\vec{\nabla} T$ lies along $x$, the LBTE is reduced to
\begin{equation}
v^{(x)}_{\mu}\nabla T \frac{\partial \bar{n}_{\mu}}{\partial T} 
+ v^{(y)}_{\mu} \frac{\partial \Delta n_{\mu}(y)}{\partial y} 
= - \frac{1}{\mathcal{V}} \sum_{\mu'} \Omega_{\mu\mu'} \Delta n_{\mu'}(y) \; .
\label{eq2}
\end{equation}
The trench case is analogous, but the translational invariance is along $y$.
We will explicitly write the equations for the case of the ribbon, with the trench obtained substituting $y\to x$ everywhere.
We also note that Eq. \ref{eq2} can be used for studying the transport properties of thin films, the trench (ribbon) geometry corresponding to the transport properties of a thin film in the transverse (longitudinal) direction.


The relaxation-time description of surface scattering, which states that the surface limits the phonon MFPs, can be derived from the equation above.
The procedure, however, requires adopting the relaxation-time approximation, in which the out-of-diagonal elements of the scattering matrix are neglected and a simple diagonal matrix $\Omega_{\mu\mu'} = \frac{1}{\tau_\mu} \delta_{\mu\mu'}$ is left, determined by the phonon lifetimes $\tau_\mu$.
The proof is straightforward, although lengthy, and was given in Ref. \citenum{carruther}; we report it in the Supplementary Note 1 for convenience.
The appealing part of this approximate theory is that one can map the solution of the finite-sized system to a bulk LBTE (much simpler to solve, since $\Delta n_\mu$ doesn't depend on space variables anymore).
Simply, one corrects the phonon lifetime adding a term $\frac{|v^x_\mu|}{L}$ through a Matthiessen sum: $\frac{1}{\tau_\mu} \to \frac{1}{\tau_\mu} + \frac{|v^x_\mu|}{L}$.
The surface correction is relevant only if the phonon lifetime is much larger than the time it takes for the phonon to travel for a distance $L$ at velocity $|v^x_\mu|$.
For very short lifetimes (high temperatures) the surface correction is negligible, but at lower temperatures, when the intrinsic phonon lifetime grows and diverges, the effective phonon lifetime is driven by the surface term, in agreement with the qualitative picture mentioned in the introduction.
In order to go beyond this approximation, modern literature \cite{fugalloPRB,fugallo-nano,broido:silicon,sparavigna:nc} often describes surface scattering by retaining off-diagonal terms for the intrinsic phonon scattering processes and adding the correction $\frac{|v^x_\mu|}{L}$ to the diagonal elements: we will refer to this approach as Surf-RTA and we will use it as a comparison, although we couldn't find a way to justify formally this approximation.

A peculiar characteristic of surface scattering is that it doesn't occur uniformly throughout the bulk, at variance with intrinsic processes such as 3-phonon or phonon-isotope scattering events. 
Therefore, surface scattering cannot be treated simply by adding a new term to the scattering operator, as one would do when considering intrinsic scatterings.
Instead, as thoroughly discussed for example in Ziman's book \cite{ziman}, the description of surface scattering must proceed by applying suitable boundary conditions (to be discussed later) to the solution of the differential equation (\ref{eq2}), which can be solved analytically without approximating the scattering matrix.
To this aim, one first rotates the LBTE in the basis of relaxons \cite{relaxons}, defined from the scattering matrix eigenvalue problem:
\begin{equation}
\frac{1}{\mathcal{V}} \sum_{\mu'} \tilde{\Omega}_{\mu\mu'}
\theta_{\mu'}^{\alpha} = \frac{1}{\tau_\alpha} \theta_\mu^{\alpha} \;,
\end{equation}
where $\tilde{\Omega}_{\mu\mu'}$ is the symmetrized version of the scattering matrix, $\alpha$ is an eigenvalue index, $\theta_{\mu}^\alpha$ is the relaxon eigenvector and $\tau_{\alpha}$ is the relaxation time of relaxon $\alpha$.
The phonon excitation number can be expressed in the basis of relaxons as $\Delta n_{\mu} = \sum_\alpha \sqrt{\bar{n}_\mu(\bar{n}_\mu+1)} f_\alpha \theta_\mu^\alpha$, where $f_\alpha$ is the relaxon population (the detailed procedure and the transformation are described in Ref. \citenum{relaxons}).
The LBTE thus becomes:
\begin{equation}
\sqrt{\frac{C}{k_B T^2}}    V^{(x)}_{\alpha}    \nabla T
+ \sum_{\beta}  V^{(y)}_{\alpha\beta}  \frac{\partial f_{\beta} }{\partial y}
= - \frac{f_{\alpha}}{ \tau_{\alpha}}  \; ,
\label{eq3}\end{equation}
where $V^{(x)}_{\alpha} = \frac{1}{\mathcal{V}} \sum_{\mu} \theta^0_{\mu} v^{x}_{\mu} \theta^\alpha_\mu $ are the relaxon velocities, $\theta^0_{\mu} = \frac{\sqrt{\bar{n}_\mu(\bar{n}_\mu+1)}\hbar\omega_\mu}{\sqrt{k_BT^2C}}$ is the linearized Bose--Einstein distribution (not an eigenvector of the scattering matrix) and $V^{(y)}_{\alpha\beta} = \frac{1}{\mathcal{V}} \sum_{\mu} \theta^\alpha_{\mu} v^{y}_{\mu} \theta^\beta_\mu $ is a matrix with physical dimension of a velocity.
Limiting the study to small deviations from equilibrium, one can search for solutions linear in $\nabla T$. 
It is convenient to make an additional variable change $f_{\alpha} = \nabla T g_{\alpha} \sqrt{\tau_{\alpha}} $, so that Eq.  \ref{eq3} becomes:
\begin{equation}
\sum_{\beta} \Lambda^{(y)}_{\alpha\beta} \frac{\partial g_{\beta} }{\partial y} + g_{\alpha} = g_{\alpha}^{\infty}
\label{eq5}
\end{equation}
where $\Lambda^{(y)}_{\alpha\beta} = \sqrt{\tau_{\alpha}} V^{(y)}_{\alpha\beta} \sqrt{\tau_{\beta}}$; $g_{\alpha}^{\infty} =  - \sqrt{\frac{C \tau_{\alpha}}{k_B T^2}} V^{(x)}_{\alpha}$ is the solution of the bulk problem (where the space dependence, i.e. the second term in the left-hand side of Eq. \ref{eq3}, disappears).
With this choice of notation, the LBTE can be seen as a set of linear differential equations coupled by the matrix $\Lambda$. Thanks to the introduction of the $g$ variables, the matrix $\Lambda$ is symmetric and also real, since both $\tau_\alpha$ and $V_{\alpha\beta}$ are real \cite{relaxons}.
Therefore, $\Lambda_{\alpha\beta}$ can be diagonalized to find a complete set of eigenvalues $\lambda_i^{(y)}$ and eigenvectors $\psi_{\alpha i}$ such that:
\begin{equation}
\sum_{\beta}\Lambda^{(y)}_{\alpha\beta} \psi_{\beta i} = \lambda_{i}^{(y)} \psi_{\alpha i}   \; ,
\end{equation}
where $i$ is the eigenvalue index.
Since the basis of $\psi_{\alpha i}$ eigenvectors is complete, the relaxon population $g$ can be expressed as:
\begin{equation}
g_{\alpha} = \sum_i \psi_{\alpha i} d_i \;,
\end{equation}
where we interpret $d_i$ as the population number of mode $\psi_{\alpha i}$.
Inverting the above relation one can think at the modes $\psi_{\alpha i}$ as a linear combination of relaxons (and thus, in turn, of phonons); thus, the $\psi_{\alpha i}$ are another set of collective phonon excitations.
Rotating the LBTE into this basis (the procedure is analogous to that followed in going from Eq. \ref{eq2} to Eq. \ref{eq3}), one obtains
\begin{equation}
\lambda_{i}^{(y)} \frac{\partial d_i}{\partial y}  + d_i = d_i^{\infty} \;,
\label{eq8}
\end{equation}
with $d_i^{\infty}$ such that $g_\alpha^{\infty}= \sum_{\alpha}  \psi_{\alpha i} d_i^{\infty}$.
So, with this choice of basis, the LBTE is reduced to a set of decoupled linear differential equations, where the populations of the collective phonon excitations $\psi_i$ can be separately solved for.
We remark in passing that we didn't derive Eq. \ref{eq8} directly from Eq. \ref{eq2}, to avoid a division of the BTE by  the phonon velocity $v_\mu$, which can be zero.
By going through Eq. \ref{eq3}, we instead can assume the correctness of the derivation provided $\tau_\alpha>0$, something that is numerically verified.

There are three possible classes of solutions for Eq. \ref{eq8}.
The first one has $\lambda_{i}^{(y)}=0$; then $d_i = d_i^{\infty}$, and the solution is equal to that of the bulk system. 
Thus the surface has no effect on the mode $i$; we interpret this case as that of a wavepacket $\psi_{\alpha i}$ traveling parallel to the surface.
We note that $\lambda_i^{(y)}=0$ is an admissible case and we show in Supplementary Note 2 that some of these solutions arise from phonons traveling parallel to the surface.
In the case $\lambda_{i}^{(y)}>0$, the population is $d_i = d_i^{\infty} + c_i e^{-y/\lambda^{(y)}_i}$, where $c_i$ is an integration constant that is determined by the boundary conditions.
If we interpret this solution as a wavepacket $\psi_{\alpha i}$ traveling in the positive direction away from the surface located at $y=-\frac{W}{2}$, there is a simple way to fix the boundary condition.
Supposing that the mode $i$ is at thermal equilibrium at the surface $y=-\frac{W}{2}$ (that is, the populations $d_i$ of mode $\psi_{\alpha i}$ must be zero, since $d_i$ are rotations of the deviation from equilibrium $\Delta n_\mu$), one can impose $d_i(y=-\frac{W}{2})=0$, finding:
\begin{equation}
d_i = d_i^{\infty} - d_i^{\infty} e^{-(y+\frac{W}{2})/\lambda^{(y)}_i} \;.
\end{equation}
We note that here we used a generalization of Casimir's boundary condition, which was originally formulated assuming phonons as heat carriers and assuming phonons to be at equilibrium at the surface (see Supplementary Note 1).
The case $\lambda_{i}^{(y)}<0$ is treated symmetrically by applying the generalized Casimir boundary condition $d_i(y=\frac{W}{2})=0$ on the opposite surface, yielding:
\begin{equation}
d_i = d_i^{\infty} - d_i^{\infty} e^{-(y-\frac{W}{2})/\lambda^{(y)}_i} \;,
\end{equation}
which fully specifies the analytical solution of the differential equation.

In summary, the exact solutions of the space-dependent LBTE are found as a sum of a bulk term and a surface correction.
Only the surface contribution depends on the position inside the ribbon, with an exponential decay depending on the distance from the surface, at a rate given by the eigenvalues of $\Lambda^{(y)}$.
The depopulation of the states $\psi_i$ is therefore localized at the surface(s) and at far distances one recovers the bulk solutions $d_i^{\infty}$.

We briefly outline a procedure to construct the boundary conditions from microscopic considerations.
Some techniques, like molecular dynamics, can estimate the phonon reflection coefficient $\mathcal{R}_\mu^{\mu'}$, i.e. the probability for a phonon $\mu$ to impact on the surface and be reflected in a state $\mu'$ traveling away from the surface.
The matrix $\mathcal{R}$ can be rotated in the $\psi_i$ basis, giving the reflection coefficient for the transition $\psi_i$ to $\psi_j$:  $\mathcal{R}_i^j = \sum_{\mu\mu'\alpha\beta}   \psi_{\alpha i} \theta_\mu^\alpha  \mathcal{R}_\mu^{\mu'} \theta_{\mu'}^\beta \psi_{\beta i} $.
A boundary condition for the surface at $y=-\frac{W}{2}$ (the other surface simply has opposite signs) is found writing a balance equation:
\begin{equation}
\delta_i = d_i\bigg(y=-\frac{W}{2}\bigg)\bigg|_{\lambda_i^{(y)} > 0} = \sum_j \mathcal{R}_i^j d_j\bigg(y=-\frac{W}{2}\bigg)\bigg|_{\lambda_j^{(y)} < 0}    \;,
\end{equation}
which states that the probability of finding a mode $i$ traveling away from the surface is equal to the sum of modes that were traveling against it and have been reflected in the mode $i$.
This condition is analogous to that written by Ziman \cite{ziman} as a starting point for the discussion of rough and smooth surfaces, but is now expressed in the $\psi_i$ basis.
Casimir's limit is a simple case in which all phonons are absorbed at the surface ($\mathcal{R}_\mu^{\mu'}=0$).
Expecting reflection coefficients to be between 0 and 1, Casimir's limit systematically underestimates the value of the boundary condition.
An atomistic simulation that would predict the values of $\mathcal{R}_\mu^{\mu'}$ goes out of the scope of this work: for simplicity we will from now on limit our investigation to the Casimir's limit, thus expecting to overestimate surface scattering.

Once the LBTE is solved, the space-dependent thermal conductivity $k$ can be readily computed.
Given the geometries under consideration, we indicate with $k$ the $xx$ component of the full thermal conductivity tensor and focus on this.
Using the definition of $k$ and rotating it to the $\psi_{\alpha i}$ basis set, we find:
\begin{align}
k(y) &=  - \frac{Q}{\nabla T}  = \frac{-1}{\mathcal{V}\nabla T} \sum_{\mu} \hbar \omega_{\mu} v_{\mu} \Delta n_{\mu} 
= -\sum_{\alpha} f_{\alpha}(y) \sqrt{k_B T^2 C} V^{(x)}_{\alpha}    
= -\sum_{\alpha} g_{\alpha}(y) \sqrt{\tau_{\alpha} k_B T^2 C} V^{(x)}_{\alpha} =  \nonumber \\
&=  \sum_{\alpha} C V_{\alpha}^{(x)} \Lambda^{(x)}_{\alpha}  -  k_BT^2 \sum_{\alpha \beta}   g_{\alpha}^{\infty} 
\bigg( \sum_{\lambda_i^{(y)}>0} \psi_{\alpha i} e^{- \frac{y+\frac{W}{2}}{ \lambda^{(y)}_{i}}} \psi^T_{i\beta} + \sum_{\lambda_i^{(y)}<0} \psi_{\alpha i} e^{-\frac{y-\frac{W}{2}}{ \lambda^{(y)}_{i}}} \psi^T_{i\beta} \bigg) 
g_{\beta}^{\infty}    \nonumber \\
&= k^{\infty} - \Delta k^{\text{surf}} (y)   \;,
\label{k_surface}\end{align}
where $\psi^T_{i\beta}$ is the matrix transpose of $\psi_{\beta i}$.
This equation is the main result of this work and shows that thermal conductivity in a crystalline nanostructure is determined by two contributions: a bulk conductivity $k^{\infty}$ that is decreased by a position-dependent surface term $\Delta k^{\text{surf}}(y)$. The latter depends on the distance from the surface through a series of exponentially decaying factors, whose decay length is determined by the eigenvalues  $\lambda_i$, which we call \emph{friction lengths} for the reasons that will be explained below.
To help the reader, we include in the supplementary material table 1 summarizing the variables used in the derivation of this result.


\begin{figure*}
\centering
\includegraphics{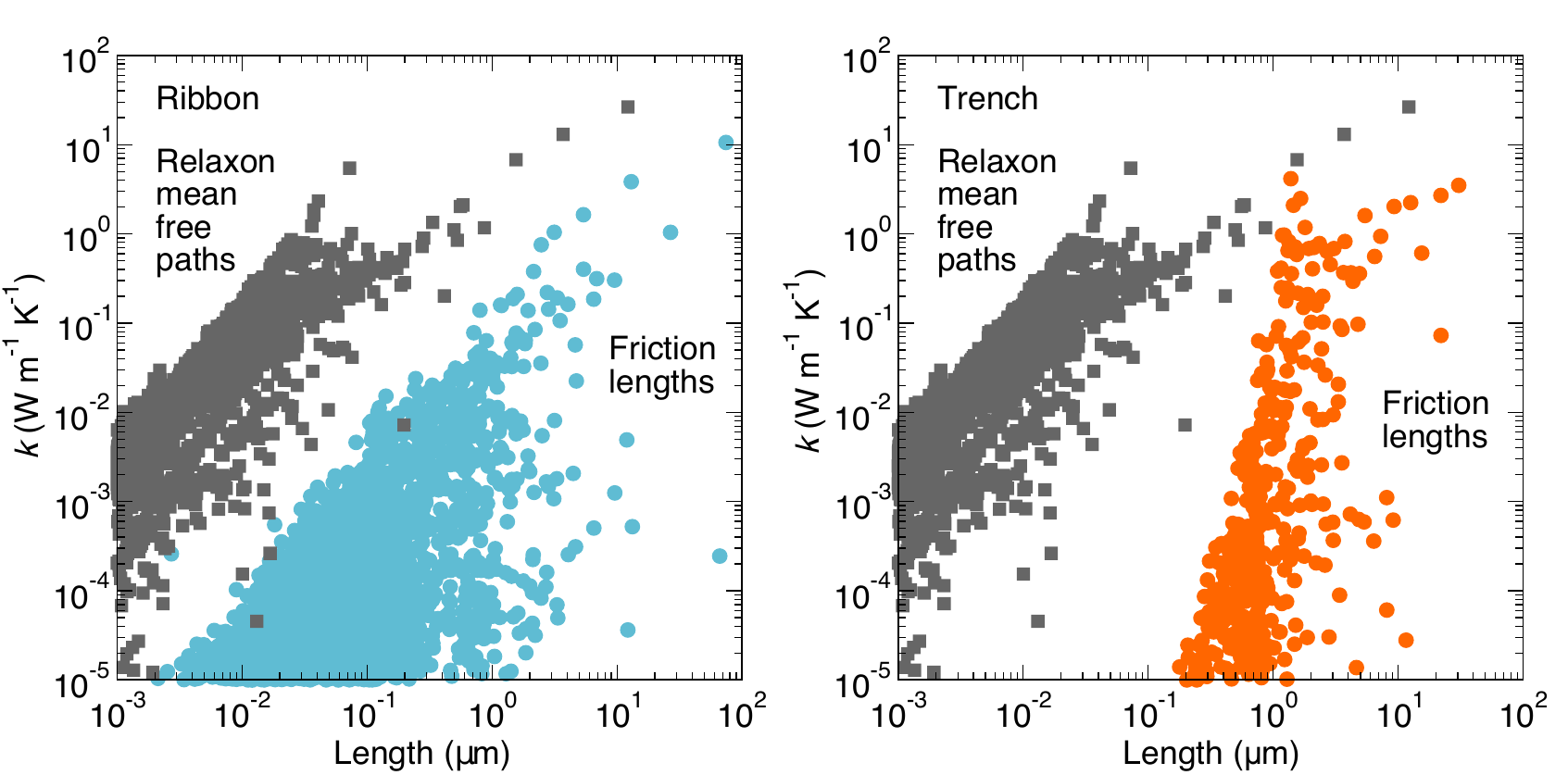}
\caption{Friction lengths (see text for definition) on the horizontal axis in a monolayer ribbon (left panel) and trench (right panel) of MoS$_2$ and their average contribution to the reduction of thermal conductivity (vertical axis) in a system of size 4 $\mu$m. 
Since every friction length has a negative contribution to the thermal conductivity (i.e. heat flux is always dissipated by surfaces), we plot the absolute value of such negative contribution. From this plot, one can infer the length scale at which nanostructuring is effective, that is, the largest friction length. We show for comparison also the mean free paths of relaxons (the heat carriers) and their (positive) contribution to bulk conductivity. We note how the relaxon mean free paths are typically shorter than the friction lengths; thus, surface scattering doesn't simply impose an upper bound to the mean free path of carriers.}
\label{fig2}
\end{figure*}

The physical consequences of Eq. \ref{k_surface} are best understood with the help of an example; here we apply it to a MoS$_2$ monolayer using the ribbon and trench geometries of Fig. \ref{fig1}, and examine the results.
When not specified otherwise, transport properties are computed at 300K, for a sample size (i.e. ribbon width or trench length) of 4 $\mu$m.
The crystal zig-zag direction lies along $x$ and for simplicity we will not consider other orientations: in general, results depend on the crystal orientation through the suitable projection of the phonon group velocities appearing in the surface contribution. The bulk contribution is isotropic in this case, as determined by the bulk crystal symmetry of MoS$_2$.
Lattice harmonic and anharmonic force constants are computed using density-functional perturbation theory \cite{baroni:rev,giannozzi:prb43,debe:prl,paulatto,lazzeri:anharmonic,baroni:prl-dfpt,gonze:2np1} as implemented in the Quantum ESPRESSO distribution \cite{qe}.
The scattering matrix $\Omega$ is built considering three-phonon and phonon-isotope scattering events \cite{fugalloPRB,relaxons} at natural abundances \cite{iupac}.
Linear-algebra operations are handled with the SCALAPACK library \cite{scalapack}, and calculations are managed using the AiiDA materials' informatics platform \cite{AiiDA}.
The remaining computational details are reported in Supplementary Note 3.

In Fig. \ref{fig2} we analyze the friction lengths $\lambda_i^{(x/y)}$ and their average contribution to $\Delta k^{\text{surf}}$. 
This latter quantity is determined by integrating the contribution from the $i$-th mode in Eq. \ref{k_surface} along the crystal size.
Fig. \ref{fig2} provides the critical information on the length scales at which nanostructuring can reduce conductivity.
In the ribbon geometry, for example, the largest friction length is about 90 $\mu$m: as a result, if the sample width is much larger than this quantity, all exponential factors are approximately zero and thus $k \approx k^{\infty}$.
Conversely, when the sample becomes narrower than 90 $\mu$m, the contribution of that mode is decreased by a factor $1/e$, and, as the sample gets smaller, more modes suppress the bulk thermal conductivity.
The trench geometry shows instead a bulk behavior above a characteristic width of 40 $\mu$m.
Each friction length has a non-negative contribution to $\Delta k^{\text{surf}}$; therefore - in this model - a finite sized sample always has a smaller conductivity than the bulk.
Numerically, we found a symmetry in the eigenvalue spectrum of $\Lambda$, such that both $\lambda$ and $-\lambda$ are valid friction lengths and contribute in an equal way to the average thermal conductivity; for this reason, we reported in Fig. \ref{fig2} only the positive part of the spectrum without loss of generality.

At this point, we are equipped to show that the widespread interpretation of surface scattering, i.e. that the finite size provides an upper bound to the carrier's MFP, does not apply.
To illustrate this point, we show in Fig. \ref{fig2} also the relaxon MFPs, i.e. the quantities associated with the distance travelled in bulk by heat before dissipation occurs.
The two quantities display very different length scales, with the distribution of relaxon MFPs shifted to length scales much smaller than the friction lengths.
Moreover, relaxon and friction lengths, despite having both the dimension of a length, operate on different objects: the relaxon MFP is associated to the collective excitation defined as the eigenvectors $\theta^\alpha_\mu$ of the scattering matrix, whereas friction lengths suppress the distance traveled by the modes $\psi_{\alpha i}$.
We must therefore elaborate further on an intuitive physical understanding of surface effects.
For completeness, we report in Supplementary Figure 1 the comparison of friction lengths with the phonon mean free paths, showing that these two quantities do not coincide as well.

\begin{figure*}
\centering
\includegraphics{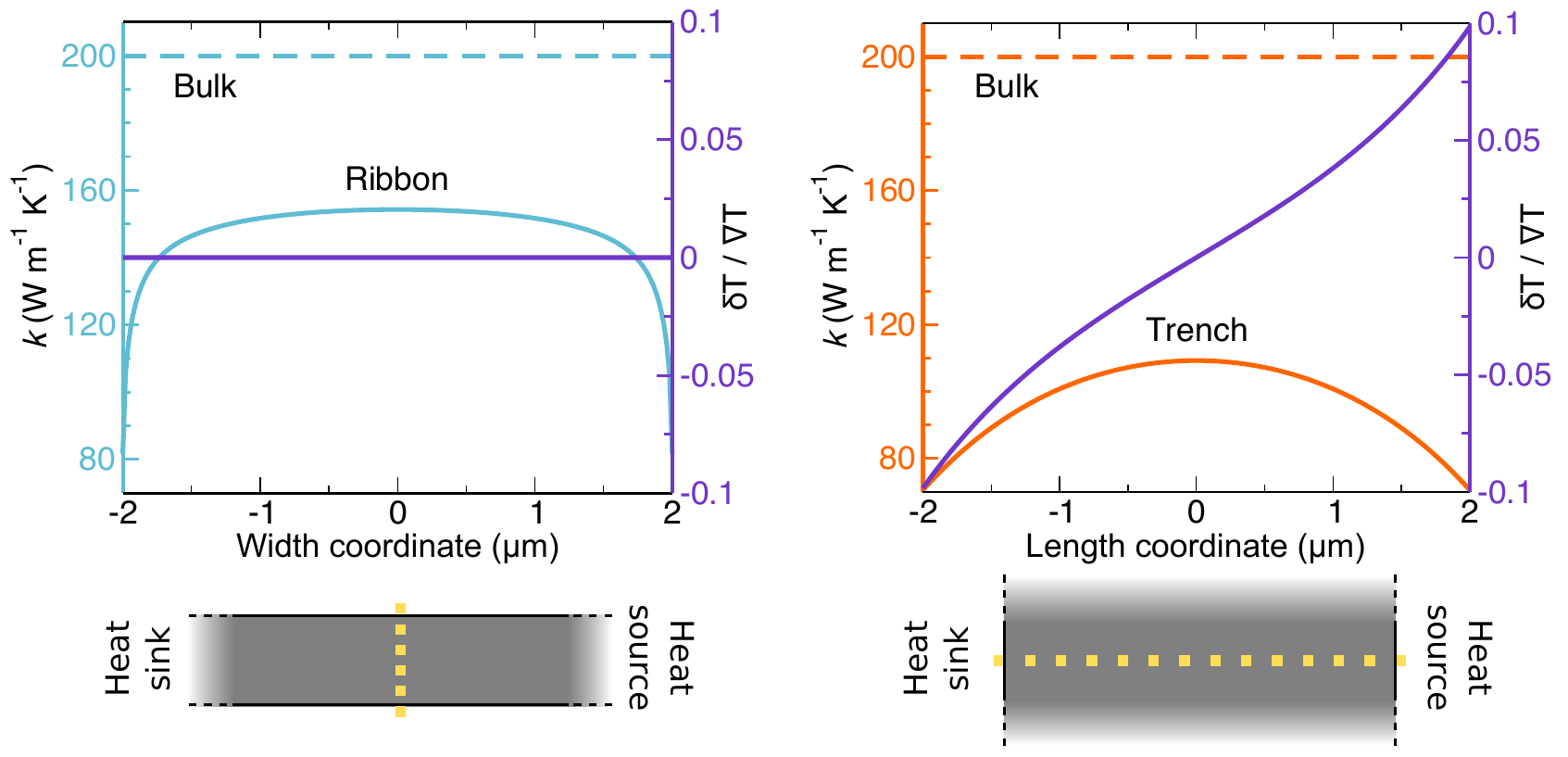}
\caption{Profiles of thermal conductivity (left axis) and temperature difference $\delta T$ (right axis) along a section of a ribbon (left panel) or trench (right panel) of monolayer MoS$_2$. The yellow dotted line in the panel below indicates the section along which quantities are plotted in the panel above. 
In presence of surfaces, thermal conductivity is not a materials intrinsic property and depends on the position inside the sample. 
In analogy to the flow of a real fluid, the thermal conductivity has a maximum at the center and minimum on the borders (the thermal conductivity at the border is 82 W/mK and 71 W/mK for the ribbon and trench respectively). Temperature changes are relative to the applied gradient of temperature $\nabla T$ and depend on the orientation of the sample.}
\label{fig3}
\end{figure*}

One can find an intuitive interpretation for this model by inspecting the profile of thermal conductivity across the crystal, as shown in Fig. \ref{fig3}.
Thermal conductivity shows a maximum at the sample center and decreases to a non-zero minimum at the two surfaces, providing a striking similarity with the Hagen-Poiseuille description of the laminar flow of a viscous fluid through a pipe.
In this latter case, the walls slow down the flux as a consequence of shear viscosity, but this effect disappears for large distances from the walls.
Here we propose to interpret $V_{\alpha\beta}$, $\Lambda_{\alpha\beta}$ and $\lambda_i$ as quantities related to the friction of the relaxon gas.
In a bulk system, the application of a temperature gradient leads to a flux of the relaxon gas that doesn't depend on the position coordinate and is limited by phonon interactions.
We estimate the room temperature bulk conductivity at 200 W/mK (for comparison, we also estimate at 124 W/mK the bulk conductivity using the single-mode relaxation time approximation, see Ref. \citenum{relaxons} for an interpretation of this approximation and its limitations).
When a surface is introduced, the heat flux is slowed down most in proximity of surfaces, where, as in a viscous fluid, a friction is induced on the relaxon gas.
Microscopically, such friction of the heat flux originates from the collision of the components $\psi_{\alpha i}$ of the relaxon gas against the surface that tend to thermalize the system.
As the distance from the surface is increased, the heat flux experiences less friction and, for sufficiently long distances, the system behaves as the bulk.
Note also that, just like their viscous counterpart in fluids, the quantities associated with friction are intrinsic properties of the bulk: the surface enters the equations only through the boundary conditions and the distance in the exponential decaying factors.
The analogy between phonons and liquids has often been used in the past \cite{nature-chen,nature-mio}, as well as in studies of electronic transport \cite{Bandurin1055,Crossno1058,Moll1061}. 
However, one should keep in mind a few key distinctions between a classical fluid and the hydrodynamic behavior of heat carriers \cite{Ashcroft}.
In fact, the particle number in a classical fluid is conserved and also momentum is conserved in particle collisions, except for scatterings at the walls: friction in such liquid slows down the particle velocities. 
Here instead, both particle number and momentum (except for normal processes) are not conserved through collisions and the particle velocity is fixed, essentially, by the phonon dispersion relations: the friction we are here considering depopulates states close to the surface.
Notwithstanding these differences, the similarities help grasping an intuitive understanding of the problem.

In Fig. \ref{fig3} we also estimate the temperature profile reconstructed from kinetic arguments.
To this aim, note that the energy deviation from equilibrium $\Delta E$ and the local temperature difference $\delta T$ are related via $\Delta E = C\delta T$, where $C$ is the specific heat.
Moreover, $\Delta E$ can be expressed in terms of relaxons \cite{relaxons}: $\Delta E = \frac{\sqrt{k_BT^2 C}}{\mathcal{V}} \sum_{\alpha \nu} f_{\alpha} \theta^0_{\alpha} \theta^\alpha_{\mu}$.
Therefore, one can reconstruct the temperature profile, noting however that $\delta T$ is, like $g_\alpha$, proportional to the applied thermal gradient $\nabla T$.
It's interesting to note that $\delta T$ behaves differently depending on the geometry and the directions chosen.
For the ribbon, the temperature profile parallel to the thermal gradient is only determined by $\nabla T$ and in the orthogonal direction $\delta T$ is simply zero (within numerical error), as shown in Fig. \ref{fig3}.
For the trench, temperature is constant in the direction orthogonal to the applied temperature gradient.
In the parallel direction, we observe in Fig. \ref{fig3} that the response temperature $\delta T$ is mostly linear, with deviations from linearity in proximity of surfaces.
The sigmoid-like behavior of the response temperature is reminescent of what can be obtained with non-equilibrium molecular dynamics simulations (see for example Refs. \citenum{PhysRevB.45.7054,PhysRevB.59.13707,PhysRevB.79.115201,C2CP42394D}).
We take the opportunity to comment that molecular dynamics simulations often use a relaxation-time description of surface scattering, as a tool to extrapolate results in the bulk limit: here we do not investigate the matter further, but simply note that Eq. \ref{k_surface} should provide guidance to improve extrapolation techniques.


\begin{figure}
\centering
\includegraphics{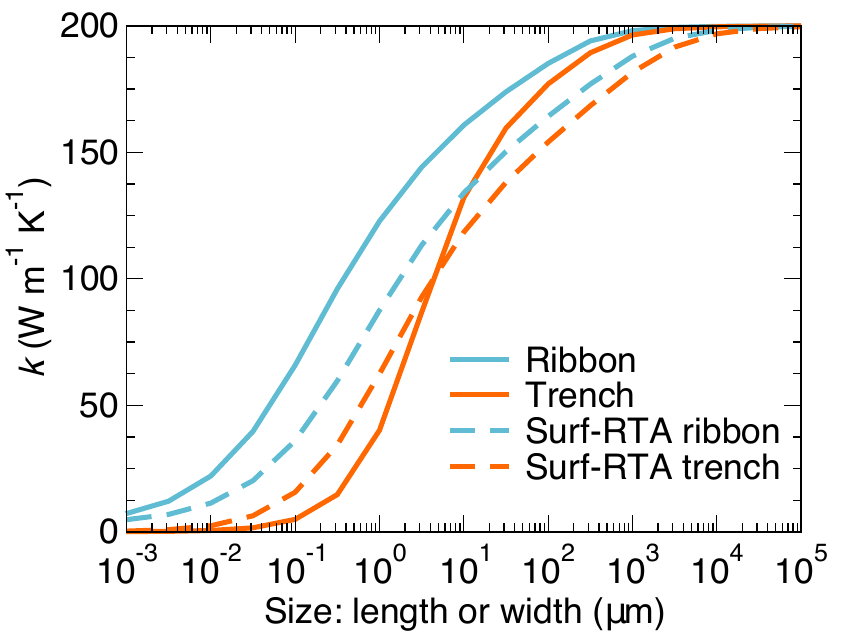}
\caption{Room temperature averaged thermal conductivity $\bar{k}$ of MoS$_2$ as a function of the sample size, i.e. width for a ribbon (in orange) and length for a trench (in blue). The results of our model (solid lines) are compared with the Surf-RTA results (dotted lines): the two theories give different results, although the qualitative behavior is similar.}
\label{fig4}
\end{figure}

In light of the previous considerations, it becomes now clear that thermal conductivity in a nanostructure cannot be considered an intrinsic property \cite{fugallo-nano,ziman}, as it depends on where it is estimated.
Nevertheless, it is convenient to tag a structure with an average thermal conductivity $\bar{k}$ (the average is performed by  analytically integrating Eq. \ref{k_surface} over the system width or length): $\bar{k}$ is a well-defined quantity, but it should not be used in a diffusion equation (such as Fourier's law) in an attempt to reconstruct a temperature profile.

In Fig. \ref{fig4}, we plot $\bar{k}$ as a function of the crystal size at room temperature and contrast our results with the commonly used Surf-RTA modeling of surface effects.
Our approach preserves the qualitative behavior of the Surf-RTA: $\bar{k}$ goes to zero in the small crystal limit and reaches bulk values for large sizes with a smooth transition in between the two cases.
Such intermediate region, which is the most relevant to estimate the impact of nanostructuring on $\bar{k}$, exhibits significant differences between the approximate Surf-RTA and our results.
The Surf-RTA systematically overestimates the impact of nanostructuring on $\bar{k}$ in the ribbon (i.e. one would need smaller sample sizes instead), while in the trench it over- (under-) estimates surface effects in the large- (small-) size limit.
Some of the values in Fig. \ref{fig4} can be compared with earlier work: for example, the Surf-RTA value of conductivity in the ribbon geometry at width 1 micron (87 W/mK) compares well with similar studies (83 W/mK in Ref. \citenum{theory:mos2}). 
It's still not possible to compare with the experimental value for MoS2 monolayer (34.5 W/mK \cite{experiment:mos2}), as the experimental setup used a geometry different from those studied here (a Corbino disk of diameter 1.2 $\mu$m).

\begin{figure}
\centering
\includegraphics{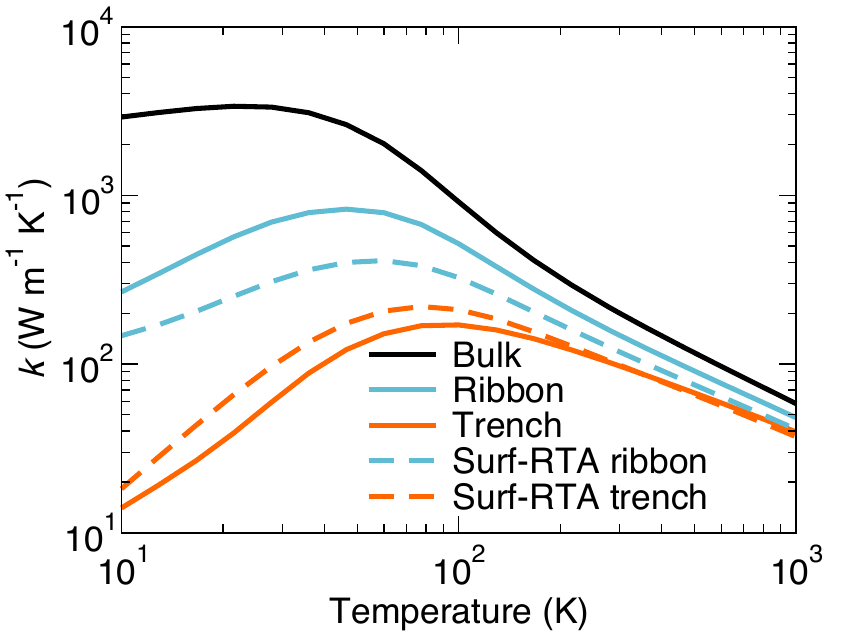}
\caption{
Averaged thermal conductivity $\bar{k}$ of a monolayer of MoS$_2$ as a function of temperature in the trench (length 4$\mu$m) and ribbon geometries (width 4$\mu$m). For comparison, we show the bulk values in black and in dotted lines the values obtained with the approximated Surf-RTA description: the largest improvement of this theory is obtained at lower temperatures, where surface effects dominate thermal transport.
}
\label{fig5}
\end{figure}

Another comparison with the Surf-RTA is shown in Fig. \ref{fig5}, where we plot the average thermal conductivity as a function of temperature, together with the bulk conductivity.
In agreement with conventional expectations, the impact of surfaces is the largest at low temperatures: here our theory has the largest departure from Surf-RTA estimates.
In fact, at smaller temperatures more Umklapp processes are frozen out \cite{ziman}, impeding heat flux dissipation by phonon scattering events and thus giving rise to more pronounced collective effects, which can only be captured by the exact solutions of the LBTE \cite{nature-mio}.
As temperature is increased, the conductivity slowly reduces the deviation from the bulk limit, although even at 1000K our estimate of $\bar{k}$ still differs from the bulk value or from the Surf-RTA estimate, even more so considering that a log-log scale is used.
Although not shown in detail, we stress that the difference between our approach and the Surf-RTA also depends on the crystal size chosen; Fig. \ref{fig4} is specific to a system of size 4 $\mu$m.
We also stress that all the concepts exposed in this model of surface scattering don't rely on the dimensionality: although we compared with the Surf-RTA in a 2D system, we expect similar trends to hold in 3D materials as well.

A final comment is due on the limitations of this model of surface scattering (which holds as well for the Surf-RTA).
By writing Eq. \ref{eq1} one implicitly assumes that lattice properties (phonon force constants and phonon interaction couplings) do not depend on position.
This assumption could be improved in the immediate neighborhood of a surface, where the force constants of the first few layers of atoms differ, in general, from the bulk.
Therefore, bulk interatomic force constants are best applied for sufficiently large crystals, such that the atomistic details of the surface can be neglected.
One could overcome this limitation by explicitly computing force constants over the entire section of the ribbon and then treat it as a 1D system (which we recommend for studying narrow nanoribbons).
In this case, an insertion method like that introduced in Ref. \citenum{bungaro:1996} could be used to study surface phonons.
As a result, the model discussed here is best used in describing crystals of mesoscopic sizes, in exchange for the loss of some atomistic details.

To summarize, we present here a novel formalism to describe the effects of surfaces on thermal conductivity, that doesn't rely on simplifications of the scattering matrix in the phonon LBTE.
We showed for the geometries studied that, starting from a relaxon basis, the LBTE can be successively reduced to a set of decoupled linear differential equations that can be solved analytically, with the appropriate choice of boundary conditions. 
We showed that the traditionally accepted picture - that a surface reduces thermal conductivity by limiting the MFP of heat carriers - does not hold anymore when the LBTE is solved exactly in terms of relaxons.
Instead, the relaxon gas is characterized by friction, that slows down the flux in proximity of the surface and exponentially decays to the bulk limit at large distances.
We applied this approach to a ribbon and a trench of monolayer MoS$_2$, showing that it predicts the profiles of temperature and thermal conductivity in the sample's interior, estimates the length scale at which nanostructuring reduces thermal conductivity and the dependence of thermal conductivity on the crystal size.
The model can be applied to any other 2D material and it can be readily extended to other geometries, such as 3D thin films or more complex geometries than the ribbon and trench discussed here.
The generality of the LBTE suggests that similar results should hold also for electronic transport, and we speculate that more analogies with fluidodynamics and hydrodynamic regimes for electrons or phonons could be explored in the future.

\section{Acknowledgement}

We gratefully acknowledge Nicolas F. Mounet for useful discussions, the Swiss National Science Foundation, the National Centre of Competence in Research MARVEL, the Max Planck-EPFL Center for Molecular Nanoscience and Technology, and the Swiss National Supercomputing Center CSCS under Project ID s580.


\clearpage

\section{Supplementary Note 1: Surf-RTA surface scattering}
Here, we would like to recall the standard modeling of surface effects on thermal conductivity within the single-mode relaxation time approximation, following the derivation of Ref. \citenum{carruther}.
This model has been used to justify an effective relaxation time that is width- (or length) dependent (see Eq. \ref{eq22} below).
Using the relaxation time approximation, the LBTE is:
\begin{equation}
\frac{\partial \bar{n}_{\mu}}{\partial T} \vec{v}_{\mu} \cdot \vec{\nabla} T + \vec{v}_{\mu} \cdot \vec{\nabla} (\Delta n_{\mu}(\vec{r})) = - \frac{\Delta n_{\mu}(\vec{r}) }{\tau_{\mu}}  \; .
\end{equation}
Both sides of the equation are diagonal in $\mu$ and there is no need to change basis set.
First, we consider a ribbon geometry as in the main text, so that $\Delta n_{\mu}(\vec{r}) = \Delta n_{\mu}(y) $ due to translational invariance along $x$, the direction of the applied thermal gradient $\nabla T$.
Limiting ourselves to small deviations from equilibrium, we look for solutions that are linear in the temperature gradient: $\Delta n_{\mu} \to \Delta n_{\mu} \nabla T$, so that the LBTE becomes:
\begin{equation}
\frac{\partial \bar{n}_{\mu}}{\partial T} v^x_{\mu} + v^y_{\mu}  \frac{\partial \Delta n_{\mu}(y)}{\partial y} = - \frac{\Delta n_{\mu}(y)}{\tau_{\mu}}  \; .
\label{eq26}\end{equation}
We now distinguish three cases.
If $v^y_{\mu}=0$, the phonon is traveling parallel to the surface of the ribbon and behaves as in the bulk (where no derivative needs to be considered).
Consider now the case $v^y_{\mu}>0$ (the case $v^y_{\mu}<0$ is analogous): we can divide the equation by $v^y_{\mu}$ and find:
\begin{equation}
 \frac{\partial \Delta n_{\mu}(y)}{\partial y}  
+ \frac{\Delta n_{\mu}(y)}{\lambda^y_{\mu}}
+ \frac{ \lambda^x_{\mu} }{\lambda^y_{\mu}} \frac{\partial \bar{n}_{\mu}}{\partial T}   = 0  \; ,
\end{equation}
where we introduced the phonon mean free paths $\lambda^{x/y}_{\mu} = v^{x/y}_\mu \tau_\mu$ along the $x$ or $y$ directions.
The solution to this differential equation is:
\begin{equation}
\Delta n_{\mu} (y) = - \lambda^x_{\mu} \frac{\partial \bar{n}_{\mu}}{\partial T}  + c_\mu e^{- y / \lambda^y_{\mu}} \;,
\end{equation}
where $c_\mu$ is an integration constant to be fixed by boundary conditions.
Next we adopt Casimir's boundary conditions, that assume the phonons to be at thermal equilibrium at the surface $y=-W/2$ from which they are traveling away, i.e. $\Delta n_\mu(y=-\frac{W}{2})=0$.
We thus find:
\begin{equation}
\Delta n_{\mu} (y) =- \lambda^x_{\mu} \frac{\partial \bar{n}_{\mu}}{\partial T} \big( 1-  e^{- (y+\frac{W}{2}) / \lambda^y_{\mu}}  \big)    \;.
\end{equation}
If phonons have negative velocity, similar arguments apply, with the boundary condition $\Delta n_\mu(y=W/2)=0$, obtaining:
\begin{equation}
\Delta n_{\mu} (y) = -\lambda^x_{\mu} \frac{\partial \bar{n}_{\mu}}{\partial T} \big( 1- e^{- (y-\frac{W}{2}) / \lambda^y_{\mu}}     \big)    \;.
\end{equation}

This is the solution of the differential equation.
However, a further modification is often adopted in the literature, by considering only the average phonon population and thus studying a bulk LBTE that mimics the effect of the surface.
To this aim, we integrate the LBTE over $y$ (Eq. \ref{eq26}) to find:
\begin{gather}\label{eq17}
\frac{\partial \bar{n}_{\mu}}{\partial T} v^x_{\mu}  = -   \big(  \frac{1}{\tau_\mu^b} + \frac{1}{\tau_{\mu}}  \big) \Delta n_{\mu}  \\
\Delta n_{\mu} = \frac{1}{W} \int_{-W/2} ^{W/2} \Delta n_{\mu}(y) dy \\
\frac{1}{\tau_\mu^b} = \frac{    \int_{-W/2} ^{W/2} v_\mu^y  \frac{\partial  \Delta n_{\mu}(y) }{\partial y }   dy    }{       \int_{-W/2} ^{W/2} \Delta n_{\mu}(y) dy }  \label{tau_surface} \;.
\end{gather}
This set of equations are equivalent to the previous LBTE, but now the equation (\ref{eq17}) to be solved appears like a bulk LBTE, with an additional surface relaxation time $\tau_{\mu}^b$ added to the scattering term.
From the solution $\Delta n_\mu(y)$, one can compute the integrals and estimate the surface relaxation time $\tau_\mu^b$.
Rather than studying the complete integral, we just mention two limiting cases.
In the limit of rare intrinsic phonon interactions (as it can happen at low temperatures), $\tau_\mu\to\infty$ and one finds:
\begin{equation}
\frac{1}{\tau_\mu^b} \to  \frac{|v_\mu^y|}{\big(\frac{W}{2}\big)} \;.
\end{equation}
Thus, in the diluted scattering case, the phonon cannot travel for a distance longer than half the sample size. Assuming instead frequent phonon interactions (at high temperatures for example), $\tau_\mu\to 0$ and one obtains:
\begin{equation}
\frac{1}{\tau_\mu^b} \to  \frac{|v_\mu^y|}{W} \;.
\label{eq21}\end{equation}
We therefore see that the phonon can cover at most the sample size.

The full integral of Eq. \ref{tau_surface} adds some complexity to this picture, but represents just an interpolation between these two limits.
If we ignore these details and simply decide to use the high temperature limit, surface effects can be studied with a bulk LBTE, with an effective relaxation time
\begin{equation}
\frac{1}{\tau_\mu^{\text{eff}}} = \frac{1}{\tau_\mu} + \frac{1}{\tau_\mu^b}  =  \frac{1}{\tau_\mu} +\frac{|v_\mu^y|}{W}  \;.
\label{eq22}
\end{equation}
We stress that the derivation of this result strictly relies on the additional simplifications of Eqs. \ref{eq17} and \ref{eq21}, in addition to the relaxation-time approximation.

During the last few years, several papers, including ours, have tried to go beyond the relaxation time approximation by keeping the full scattering matrix $\Omega_{\mu\mu'}$, where the surface scattering has been treated by adding the surface relaxation-time to the diagonal matrix element of $\Omega_{\mu\mu'}$:
\begin{equation}
\Omega_{\mu\mu'}^{\text{eff}}  = \Omega_{\mu\mu'}  + \frac{ 1 }{\tau_{\mu}^b} \delta_{\mu\mu'} \;.
\end{equation}
Such approach, however, appears unjustified and therefore surface scattering needs to be tackled as shown in the main text.

\section*{Supplementary Note 2: solutions traveling parallel to the surface}
In this note we show that states with $\lambda_i=0$ exist.
We take, like in the main text, the case of a ribbon geometry, for which surfaces are parallel to the $x$ axis.
The starting hypothesis is that some phonon states $\mu$ exist, such that they travel parallel to the surface, i.e. $v^y_\mu=0$. 
We note that the matrix $V^{(y)}_{\alpha\beta}$ is non-diagonal in the relaxon basis, but is diagonal in the phonon basis; then, its eigenvalues are equal to the phonon group velocities $v^y_\mu$.
Therefore, the matrix $V^{(y)}_{\alpha\beta}$ has some eigenvectors with zero eigenvalues.
Performing the scaling of Eq. 5, one can prove that the zero-valued eigenvectors of $V^{(y)}_{\alpha\beta}$ are also eigenvectors of $\Lambda^{(y)}_{\alpha\beta} = \sqrt{\tau_\alpha} V^{(y)}_{\alpha\beta} \sqrt{\tau_\beta}$ with zero eigenvalues.
Therefore, all phonons traveling parallel to the surface are populated as in the bulk case.

We conclude this Note with two remarks.
First, this proof doesn't rule out the existence of other modes with $\lambda_i=0$; therefore, other modes may exist that do not scatter on the surface.
Finally, the eigenvectors of $\Lambda^{(y)}_{\alpha\beta}$ are in general different from phonons.
Only the simple case of zero eigenvalues can be traced back to properties of single phonons; in general, the eigenvectors will have contributions from multiple phonon states.

\section*{Supplementary Note 3: Methods}
\subsection*{First-principles simulations}
Density-functional theory calculations have been performed with the Quantum ESPRESSO distribution \cite{qe}, using the local-density approximation and norm-conserving pseudopotentials from the PSLibrary \cite{pslib} with a plane-wave cutoff of 100 Ry.
MoS$_2$ is simulated in a slab geometry using an optimized lattice parameter $a=5.896$ Bohr and a cell height of 28 Bohr.
The periodic boundary condition does not truncate the interaction between parallel replicas of MoS$_2$ flakes, we thus expect that optical phonon frequencies are affected by long-ranged electrostatic interactions \cite{sohier:2017}. 
The Brillouin zone is integrated with a Gamma-centered Monkhorst-Pack mesh of 24$\times$24$\times$1.
Second and third order force constants are computed on meshes of 16$\times$16$\times$1 and 6$\times$6$\times$1 points, and are later Fourier-interpolated on finer meshes.
The bulk thermal conductivity of MoS$_2$ reported by coworkers and us in Ref. \citenum{nature-mio} differs from that of this work due to a problem (now corrected) in the computation of the non-linear core-correction of the molybdenum pseudopotential in the computation of third derivatives \cite{paulatto}.

\subsection*{Thermal conductivity simulations}
The scattering matrix $\tilde{\Omega}$ includes 3-phonon interactions and harmonic isotopic scattering \cite{fugalloPRB,marzari:prl} at natural abundances \cite{iupac} (94.93\% $^{32}$S, 0.76\% $^{33}$S, 4.29\% $^{34}$S, 0.02\% $^{36}$S for sulfur; 14.84\% $^{92}$Mo, 9.25\% $^{94}$Mo, 15.92\% $^{95}$Mo, 16.68\% $^{96}$Mo, 9.55\% $^{97}$Mo, 24.13\% $^{98}$Mo, 9.63\% $^{100}$Mo for molybdenum.
The scattering matrix is built using a Gaussian smearing of 5 cm$^{-1}$ and a mesh of 120$\times$120$\times$1 points for integrating the Brillouin zone.
The scattering matrix and the friction matrix are diagonalised exactly using the routine PDSYEV of the Scalapack library \cite{scalapack}.
The simulation cell is renormalized using the interlayer distance of bulk MoS$_2$ ($c/a=1.945$), in order to have a thermal conductivity comparable with the 3D counterpart.
Calculations have been managed using the AiiDA materials' informatics platform \cite{AiiDA}

\clearpage
\section*{Supplementary Figure 1}

\begin{figure}[h]
\centering
\includegraphics{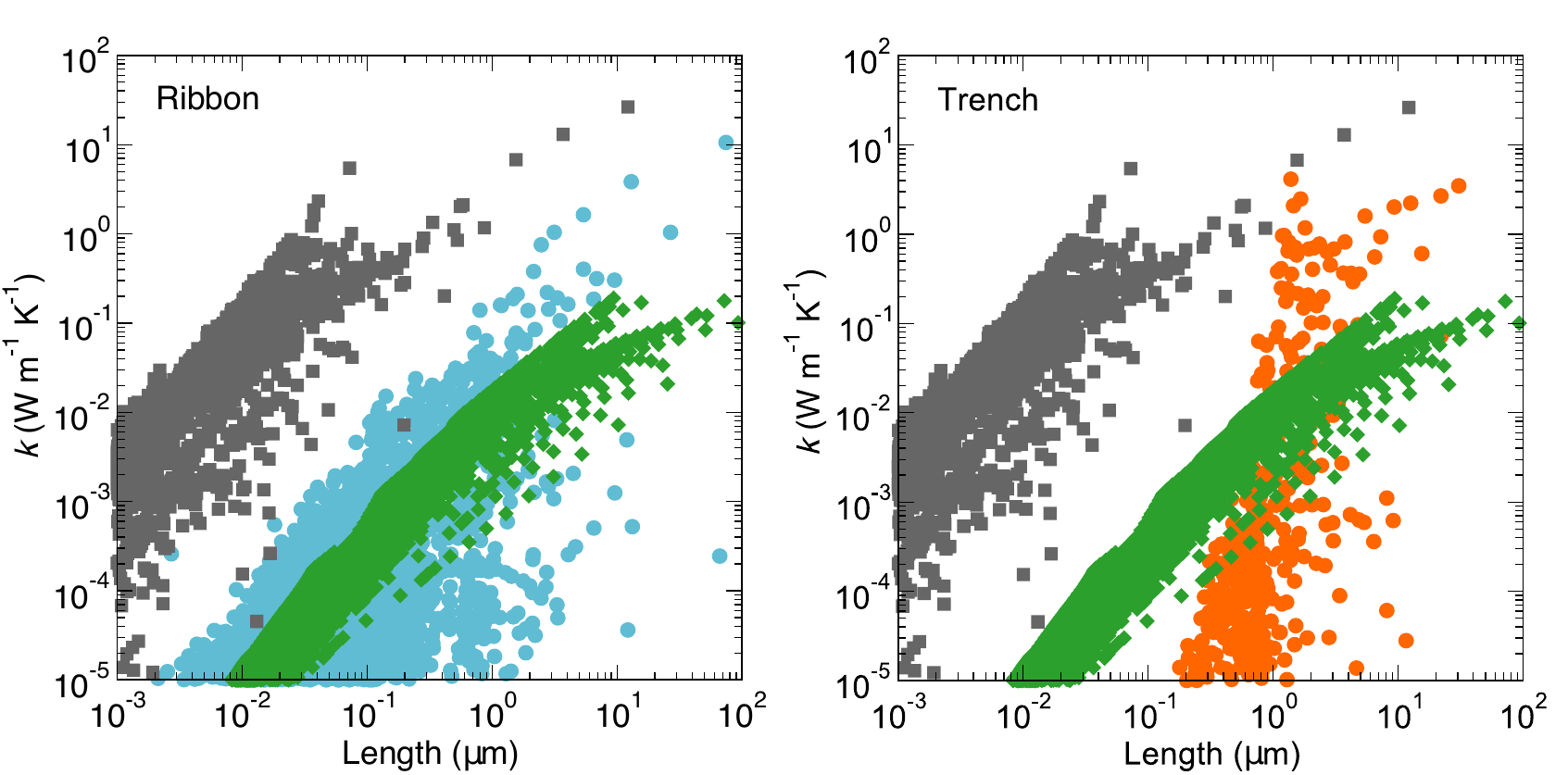}
\caption{We here overlay the phonon mean free paths to Figure 2 of the main text. The friction lengths (horizontal axis) in a monolayer of MoS$_2$ in the ribbon (blue) and trench (orange) geometry are shown together with their contribution to thermal conductivity reduction (vertical axis). We contrast this quantity with both the contribution to the thermal conductivity from relaxons as a function of their mean free paths (gray) and the contribution to the thermal conductivity from phonons (green) as a function of the phonon mean free path (i.e. within the single-mode relaxation time approximation).
We note that the differences between friction lengths and phonon mean free paths reflect the differences between the surface effect estimated with the Surf-RTA and this work reported in Fig. 4 of the main text.
In the ribbon geometry the phonon mean free paths are typically longer than the friction lengths, leading to the overestimation of the surface effect shown in Fig. 4. In the trench case, we can distinguish two regions (approximately above/below 1 micron), where phonon mean free paths are longer/shorter than friction lengths: above/below this threshold, the Surf-RTA over/under-estimates the thermal conductivity reduction.
}
\label{fig1}
\end{figure}

\clearpage
\section*{Supplementary Table 1}

\begin{table}[h]
    \begin{tabular}{ll}
    \toprule
	$x$, $y$				&	Cartesian coordinates	\\
	$\nabla T$, $k_BT$		&	Temperature gradient, thermal energy	\\
	$\mu$				&	Phonon state	\\
	$\vec{v}_\mu$			&	Phonon group velocity	\\
	$\bar{n}_\mu$			&	Bose--Einstein distribution function	\\
	$C$					&	Lattice specific heat at constant volume	\\
	$\Omega_{\mu\mu'}$	&	Phonon scattering matrix	\\
	$\frac{1}{\mathcal{V}}\sum_{\mu'}\Omega_{\mu\mu'} \theta_{\mu'}^\alpha = \frac{1}{\tau_\alpha} \theta_\mu^\alpha$	&	Relaxon eigenvalue problem	\\
	$\theta_\mu^0 = \frac{\sqrt{\bar{n}_\mu(\bar{n}_\mu+1)}\hbar \omega_\mu}{\sqrt{k_BT^2C}}$	&	Linearized Bose--Einstein distribution 	\\
	$\vec{V}_\alpha = \sum_\mu \theta_\mu^0 \vec{v}_\mu\theta_\mu^\alpha $			&	Relaxon velocity	\\
	$\vec{V}_{\alpha\beta} = \sum_\mu \theta_\mu^\alpha \vec{v}_\mu\theta_\mu^\beta $	&	Relaxon velocity matrix	\\
	$\Lambda^{(y)}_{\alpha\beta} = \sqrt{\tau_\alpha} V^{(y)}_{\alpha\beta} \sqrt{\tau_\beta}$	&	Friction length matrix	\\
	$\sum_{\beta}\Lambda^{(y)}_{\alpha\beta} \psi_{\beta i} = \lambda_{i}^{(y)} \psi_{\alpha i}$	&	Friction eigenvalue problem	\\
	$\Delta n_\mu = n_\mu - \bar{n}_\mu$							&	Change of phonon occupation number	\\
	$f_\alpha$: $\Delta n_\mu=\sum_\alpha\sqrt{\bar{n}_\mu(\bar{n}_\mu+1)} f_\alpha \theta_\mu^\alpha$	&	Relaxon occupation number	\\
	$g_\alpha = f_\alpha / (\sqrt{\tau_\alpha} \nabla T)$					&	Scaled relaxon occupation number	\\
	$d_i$: $g_\alpha=\sum_i \psi_{\alpha i} d_i$							&	Occupation number of $\psi$ modes	\\
	$g_{\alpha}^{\infty} =  - \sqrt{\frac{C \tau_{\alpha}}{k_B T^2}} V^{(x)}_{\alpha}$	&	Bulk scaled relaxon occupation number	\\
	$d_i^{\infty}$: $g_\alpha^{\infty}= \sum_{\alpha}  \psi_{\alpha i} d_i^{\infty}$	&	Bulk occupation number of $\psi$ modes	\\
	$k^{\infty}$													&	Bulk thermal conductivity	\\
	$k(x,y)$														&	Local thermal conductivity	\\
	$\bar{k}$														&	Averaged thermal conductivity	\\
	\bottomrule
    \end{tabular}
    \caption{A summary of the variables used in this work.}
\end{table}

\clearpage

\bibliography{bibliography}

\end{document}